\documentclass[prb,aps,twocolumn]{revtex4}

\usepackage{amssymb}
\usepackage{amsmath}
\usepackage{graphicx}
\usepackage{dcolumn}
\usepackage{epsf}
\usepackage{bm}
\begin{document}

\title{The Interaction in the Macroscopically Ordered Exciton State}

\author{Sen Yang}
\author{A.V. Mintsev}
\altaffiliation[Present address: ]{Institute of Solid State Physics,
Russian Academy of Sciences, 142432 Chernogolovka, Russia}
\author{A.T. Hammack}
\author{L.V. Butov}
 \affiliation{Department of Physics, University of California at San Diego, La Jolla, CA 92093-0319}
\author{A.C. Gossard}
\affiliation{Materials Department, University of California at Santa
Barbara, Santa Barbara, CA 93106-5050}

\date{\today}

\begin{abstract}
The macroscopically ordered exciton state (MOES) - a periodic array
of beads with spatial order on a macroscopic length - appears in the
external exciton rings at low temperatures below a few Kelvin. Here,
we report on the experimental study of the interaction in the MOES. The
exciton PL energy varies in concert with the intensity along the
circumference of the ring, with the largest energy found in the
brightest regions. This shows that the MOES is characterized by the
repulsive interaction and is not driven by the attractive
interaction.
\end{abstract}


\maketitle

Spatial photoluminescence (PL) patterns have been observed recently
in structures with coupled \cite{Butov02,Snoke02,Butov04} and single
\cite{Rapaport} quantum wells. The pattern features include the
inner exciton rings \cite{Butov02}, the external exciton rings
\cite{Butov02,Snoke02,Butov04,Rapaport}, the localized bright spots
(LBS) \cite{Butov02,Butov04,Butov02a,Lai04}, and the macroscopically ordered
exciton state (MOES) - a periodic array of beads with spatial order
on a macroscopic length \cite{Butov02,Butov04}. The inner and outer
exciton rings and LBS are observed up to high temperatures and are
classical phenomena. Their origin has been identified: the inner
ring has been explained in terms of nonradiative exciton transport
and cooling \cite{Ivanov06} and the external rings and LBS have been
explained in terms of macroscopic in-plane charge separation and
exciton formation at the interface of the electron- and hole-rich
regions \cite{Butov04,Rapaport}. On the contrary, the MOES is a
low-temperature phenomenon. The MOES appears in the external rings at
low temperatures below a few Kelvin \cite{Butov02,Butov04}. Because
of their long lifetime and high cooling rate, indirect excitons in
coupled quantum wells (CQW), Fig. 1a, form a system where a cold and
dense exciton gas can be created \cite{Butov:2004}. Research to understand
the origin of the MOES is in progress.

Spontaneous macroscopic ordering is a general phenomenon in pattern
formation. For instance, the MOES is characterized by a 1D spatial
modulation and periodic 1D patterns are observed in a variety of
both quantum and classical systems. The examples include the soliton
trains in atom Bose-Einstein condensates (BEC) \cite{Hulet}, Taylor
vortices in liquids \cite{Taylor}, Turing instabilities in
reaction-diffusion systems \cite{Castets} and bacteria colonies
\cite{Levine}, and gravitational instabilities in cosmological
systems \cite{Chandrasekhar}. All of these ordered states originate
from an instability, which is generated by a positive feedback to
density modulation. A particular mechanism of the positive feedback,
which is responsible for the soliton train formation \cite{Hulet}
and gravitational instability \cite{Chandrasekhar}, is an attractive
interaction: In the experiments on atom BEC, the stripe of atomic
BEC was homogeneous in the case of repulsive interaction and,
conversely, was fragmented to the periodic soliton train in the case
of attractive interaction due to the modulational instability
\cite{Hulet}; Gravitational instability results in the fragmentation
of gaseous slabs and filaments to a periodic array of high-density
clumps that is a step towards the formation of stars
\cite{Chandrasekhar}.

In this paper, we address an issue of interaction in MOES. An
attractive interaction could in principle lead to the density
modulation in the exciton ring, i.e. to the MOES formation. However,
our experimental results indicate that the interaction is repulsive
in MOES. Therefore, the interaction cannot be responsible for the
MOES formation and, on the contrary, acts against the density
modulation.

\begin{figure}
\centerline{ \includegraphics[width=0.8\columnwidth]{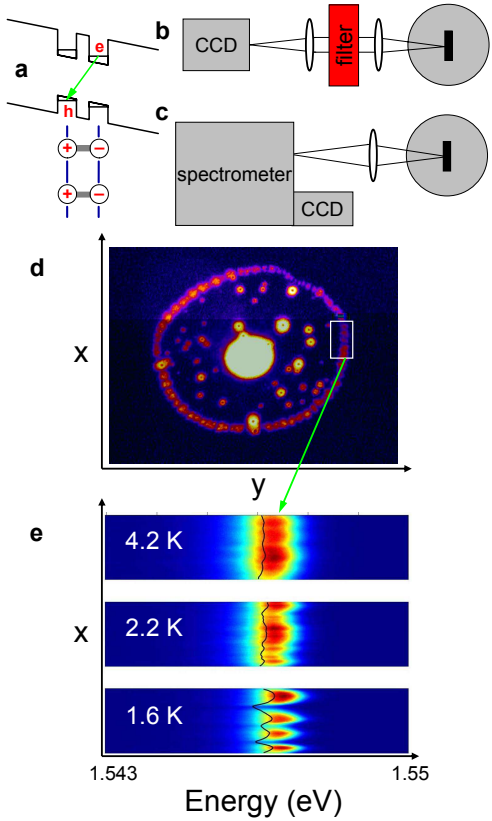} }
\caption{(a) Energy band diagram of the CQW structure; $e$ is
electron, $h$ - hole. (b) Scheme of the experimental setup for
imaging PL signals in $x-y$ coordinates (b) and $E-x$ coordinates
(c). (d) PL image of indirect excitons in $x-y$ coordinates at
$T=1.6$ K. The area of view is $365 \times 275 \,\mu$m. (e) Spectrally
resolved PL image of indirect excitons in $E-x$ coordinates for the
ring segment, which is marked by a rectangle in (d), at $T=1.6,
2.2$, and 4.2 K. The length of view (vertical axis) is $50 \,\mu$m for
each image. The black lines show variation of the PL energy of the
indirect excitons along the circumference of the ring. $V_g=1.211$ V
and $P_{ex}=0.25$ mW for the data.} \label{1}
\end{figure}

An electric-field-tunable $n^+-i-n^+$ GaAs/(Al,Ga)As CQW structure
was grown by molecular beam epitaxy. The $i$-region consists of two
8~nm GaAs QWs separated by 4~nm Al$_{0.33}$Ga$_{0.67}$As barrier and
surrounded by two 200~nm Al$_{0.33}$Ga$_{0.67}$As barrier layers.
The $n^+$-layers are Si-doped GaAs with $N_{\rm Si}=5 \times
10^{17}$~cm$^{-3}$. The electric field in the $z$-direction is
created by the external gate voltage $V_g$ applied between
$n^+$-layers. Details on the CQW structures can be found in Ref.
\cite{Butov:2004}. The indirect excitons in the CQW structure are
formed from electrons and holes confined to different QWs (Fig. 1a).
The carriers were excited by cw HeNe laser. The small disorder in
the CQW is indicated by the PL linewidth of about 1~meV. The
experiments were performed in a He$^4$ optical cryostat.

The spatial $x-y$ photoluminescence (PL) pattern was measured by a
liquid-nitrogen-cooled CCD camera after spectral selection by an
$800\pm5$ nm interference filter chosen to match the indirect exciton
energy. As a result, the low-energy bulk emission, higher-energy
direct exciton emission, and scattered laser light were effectively
removed and the indirect exciton PL emission intensity was directly
visualized in $x-y$ spatial coordinates. The scheme of the
experiment is shown in Fig.~\ref{1}b. Fig.~\ref{1}d shows the
indirect exciton PL pattern at $T=1.6$ K and excitation power
$P_{ex}=0.25$ mW. The laser excitation spot is in the center of the
image. The external exciton ring, LBS, and MOES are clearly seen in
the image (the inner exciton ring is not seen for the presented
contrast and its images can be found in Ref. \cite{Ivanov06}).

In addition, we measured the exciton PL energy profiles along the
ring using the experimental scheme shown in Fig.~\ref{1}c: a segment of
the ring was projected on the spectrometer slit and the image was
dispersed by a spectrometer without spectral selection by an
interference filter. The selected part of the external ring was
parallel to the spectrometer entrance slit and the slit width
($0.1$mm) was small enough to get a high spectral resolution
($0.1$nm) and spatial resolution in the $x$-direction ($2\mu$m). The
measured images of the PL signal in the \emph{energy-coordinate}
plane are presented in Fig.~\ref{1}e for different temperatures.

Fig.~\ref{1}e shows the indirect exciton energy profile along the
selected part of the ring. The black lines show variation of the
average PL energy $E_{PL}(x)=\frac{\int\int
E(x,y,\lambda)I(x,y,\lambda)dyd\lambda}{\int\int
I(x,y,\lambda)dyd\lambda}$ of the indirect excitons along the
circumference of the ring. The indirect exciton PL intensity
modulation increases drastically with reducing temperature, which
indicates formation of the MOES. This is consistent with the earlier
studies \cite{Butov02,Butov04}. The measurements presented in
Fig.~\ref{1}e show that the increase of PL intensity modulation with
reducing temperature is accompanied by the increase of PL energy
modulation.

\begin{figure}
\centerline{ \epsfxsize=0.8\columnwidth \epsffile{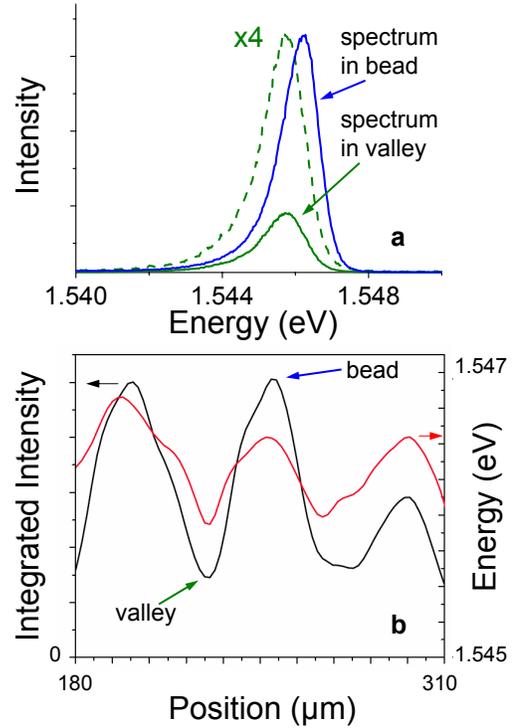} }
\caption{(a) The indirect exciton PL spectrum in the bead center
(bold blue line) and in between two beads (thin green line, dashed
green line shows the same spectrum multiplied by a factor of 4 for
comparison). (b) Variation of the PL energy and intensity of the
indirect excitons along the circumference of the ring. $T=1.6$ K,
$V_g=1.30$ V and $P_{ex}= 8.6$mW for the data. The indirect exciton
energy increases with increasing density indicating repulsive
interaction in the regime of MOES. } \label{2}
\end{figure}

The variations of the spectrally integrated PL intensity
$I_{PL}(x)=\int\int I(x,y,\lambda)dyd\lambda$ and average energy
$E_{PL}(x)$ of the indirect excitons along the circumference of the
ring are shown in Fig.~\ref{2}b. The PL energy varies in concert
with the intensity along the circumference of the ring, with the
largest energy found in the brightest regions. The corresponding
spectra in a bead center and in a valley between two beads are shown
on Fig.~\ref{2}a. The indirect exciton PL intensity and energy are
higher in the bead center.

The CQW geometry \cite{Butov:2004} is engineered so that the
interaction between excitons is repulsive: Indirect excitons, formed
from electrons and holes confined to different QWs, behave as
dipoles oriented perpendicular to the plane and an exciton or
electron-hole density increase causes an enhancement of energy
\cite{Yoshioka,Zhu,Lozovik}. The repulsive character of interaction was
evident in earlier experiments as a positive line shift with
increasing density \cite{Butov1994}. Observation of the same
behavior in the regime of the MOES (Fig.~\ref{2}) shows that the
MOES is characterized by the repulsive interaction and is not driven
by the attractive interaction.

\begin{figure}
\centerline{ \epsfxsize=0.8\columnwidth \epsffile{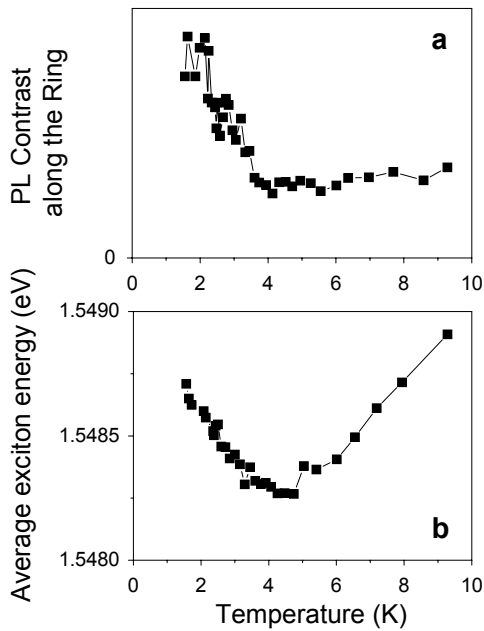} }
\caption{(a) Exciton density Fourier transform peak height at the
fragmentation period and (b) spatially averaged energy of indirect
excitons in the ring vs. temperature. $V_g=1.15 V$ and $P_{ex}= 118
\mu$W for the data.} \label{3}
\end{figure}

We have also addressed an issue whether formation of MOES lowers the
total energy of the exciton system in the ring. For this purpose, we
measured the spatially average energy of indirect excitons in the
external ring $E_{avg}=\frac{\int\int\int
E(x,y,\lambda)I(x,y,\lambda)dxdyd\lambda}{\int\int\int
I(x,y,\lambda)dxdyd\lambda}$ as a function of temperature, see
Fig.~\ref{3}b. Exciton density Fourier transform peak height at the
fragmentation period, which characterized the density modulation
contrast, is also presented in Fig.~\ref{3}a for comparison.
Fig.~\ref{3}b shows that the average energy of the indirect excitons
in the ring depends non monotonically on temperature with a
transition around $T_{tr} \sim 4$ K: with reducing temperature, the
energy reduces above $T_{tr}$ and increases below $T_{tr}$. Note that the average
indirect exciton PL energy exhibits a transition simultaneously with
the MOES onset at ca. 4 K, compare Fig.~\ref{3}a and Fig.~\ref{3}b.

Fig.~\ref{3} indicates that MOES formation is accompanied by an
increase of the total energy of excitons in the ring and, therefore,
the density modulation is not caused by lowering the total energy of
the system. Note that such a behavior is not unusual for pattern formations in
quasi-equilibrium systems and, in particular, is consistent with a
model \cite{Levitov} showing that a spatially modulated exciton
state can result from a nonlinear density dependence of the exciton
formation rate in the ring.

To conclude, the exciton PL energy has been found to vary in concert
with the intensity along the circumference of the ring for the
macroscopically ordered exciton state. This shows that MOES is
characterized by the repulsive interaction and eliminates attractive
interaction as a possible mechanism that could lead to the MOES
formation.


This work is supported by NSF grant DMR-0606543 and ARO grant
W911NF-05-1-0527. We thank K.L. Campman for growing the high quality
samples, J. Keeling, L.S. Levitov, and B.D. Simons for discussions,
G.O. Andreev and E. Shipton for help in preparing the experiment.


\end{document}